\documentclass{epl}
\title{Persistence at the onset of spatiotemporal
intermittency in coupled map lattices}
\shorttitle{Persistence in Coupled Map Lattices}
\author{Gautam I. Menon\thanks{E-mail:\email{menon@imsc.ernet.in}},
Sudeshna Sinha\thanks{E-mail:\email{sudeshna@imsc.ernet.in}}
\and Purusattam Ray\thanks{E-mail:\email{ray@imsc.ernet.in}}}
\shortauthor{G.I. Menon, S. Sinha \and P. Ray}
\institute{The Institute of Mathematical Sciences, C.I.T Campus,
           Taramani, Chennai 600 113, India
}

\begin{document}

\maketitle

\begin{abstract}
We study persistence in coupled circle
map lattices at the onset of spatiotemporal
intermittency, an onset which marks a continuous
transition, in the universality class of directed
percolation, to a unique absorbing state.
We obtain a local persistence exponent of
$\theta_l = 1.49 \pm 0.02$ at this transition,
a value which closely matches values for $\theta_l$ obtained in 
stochastic models of directed percolation.
This result constitutes suggestive
evidence for the universality of persistence
exponents at the directed percolation transition.
Given that many experimental systems are modelled
accurately by coupled map lattices, experimental
measurements of this persistence exponent may
be feasible.
\end{abstract}

The study of extended dynamical systems is
relevant to the understanding of many
phenomena in condensed matter physics, such as
pattern formation and non-equilibrium phase
transitions in coupled chemical reactions,
charge density waves and Josephson junction
arrays\cite{joseph}. A simple model which
captures much of the underlying complexity
of these systems is the coupled map lattice,
defined as a collection of elements on a
lattice which locally exhibit chaotic dynamics,
together with a diffusive local coupling between
these elements\cite{Kaneko}. Coupled map lattices exhibit
a remarkable variety of behaviour, ranging
from periodic spatio-temporal structure to
intermittency and chaos.

Spatially extended dynamical systems exhibiting spatio-temporal intermittency
possess both laminar and turbulent phases. The
laminar phase is characterized by periodic
or even weakly chaotic dynamics, while no
spatio-temporally regular structure exists in the
turbulent regime.  Spatio-temporal intermittency
refers to the properties of the state which
marks the transition between
laminar and turbulent phases. A typical
spatial pattern arising in spatio-temporal
intermittency consists of fluctuating domains
of laminar regions interspersed amidst turbulent ones.
Such intermittency is often
the precursor of fully developed
chaos\cite{Kaneko}.

An initially laminar site becomes turbulent
only if at least one of its neighbours was
turbulent at the previous time.  A turbulent
site can either relax spontaneously to a laminar
state or contaminate its neighbours at the next
time step\cite{type1}. In some regimes of parameter
space,  turbulent states percolate
in space-time, leading to an active phase.
In other regimes,
turbulent states die out, following which
the system is confined to the laminar (equivalently,
inactive) state for all time.  This laminar
state is thus a unique absorbing state.

It has been conjectured that critical exponents
associated with the onset of spatiotemporal
intermittency belong generically to the
universality class of directed percolation\cite{Pomeau}.
Underlying this conjectured correspondence is
the following more general question:
Are transitions in deterministic chaotic
systems governed by the universality classes
of stochastic models\cite{Grassberger}? To address this issue,
we study persistence and persistence 
exponents in 
coupled map lattices at the onset of
spatiotemporal intermittency.
Persistence exponents for stochastic
systems have attracted attention
recently as they appear to be a new class of
exponents, not derivable in general from the
other static and dynamic exponents\cite{pers0}.
Systems which exhibit a non-trivial persistence exponent include 
Ising and Potts models in one and higher dimensions,
the diffusion equation, the random velocity and random 
acceleration problems and several models for interface growth
\cite{satya}.

Our principal result is the following: We
find that the local persistence exponent $\theta_l$
at the onset of spatiotemporal intermittency
exhibits a remarkable universality at the
critical points 
we have examined. The
value of the persistence exponent is $\theta_l =
1.49 \pm 0.02$, in close agreement with values
obtained for stochastic models with discrete
states in one dimension, at the directed percolation 
transition\cite{hinrich,hinrich_review,albano}.
Our model, however, differs from these earlier
models in that it is {\em not} stochastic
but deterministic. Another important difference
is that the local variable in our case is
a {\em continuum} variable, unlike in the models
studied earlier.

This result is unusual for the following
reason: The persistence exponent is known to
be far less universal than the other three
exponents which typically characterize scaling
behaviour in interacting stochastic systems, the
static exponents 
$\eta$ and $\nu$ and the dynamical exponent $z$. The
reason is that persistence probes the full, in
general {\em non-Markovian}, time evolution of
a local fluctuating variable, such as a spin or
density field, from its initial state\cite{satya}. Knowing
the asymptotic properties of the evolution kernel
governing this time evolution is insufficient to
obtain the persistence exponent, although in many
cases progress can be made through controlled
expansions about Markov processes\cite{satya}. Thus,
such universality of the persistence
exponent at the directed percolation transition,
in particular for systems with deterministic
chaos, is {\it a priori\/} unexpected.

We study the dynamics of a 1-dimensional
coupled map lattice, with on-site circle
maps coupled diffusively to nearest neighbours
\cite{Kaneko, gauri}:
\begin{equation}
\label{one}
x_{i,t+1} = f(x_{i,t}) + \frac{\epsilon}{2}
(f(x_{i-1,t})+f(x_{i+1,t}) - 2 f(x_{i,t})) \ \ \ {\rm
mod} \ \ 1 
\end{equation}
where $t$ is the discrete time index, and $i$ is the site index: $i=1,
\dots L $, with $L$ the system size. The parameter $\epsilon$ measures
the strength of the diffusive coupling between site $i$ and its two
neighbours. The on-site map is chosen to be
\begin{equation}
\label{two}
f(x) = x + \omega -\frac{k}{2 \pi} \sin(2 \pi x) 
\end{equation}
where the parameter $k$ denotes the nonlinearity. All sites
are updated in parallel. 

\begin{figure}
\twoimages[scale=0.875]{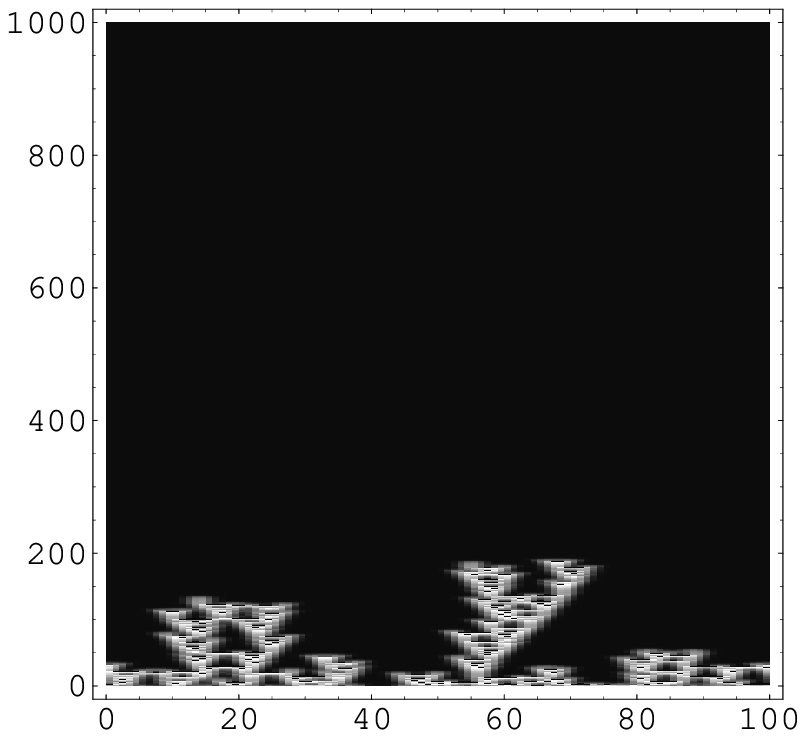}{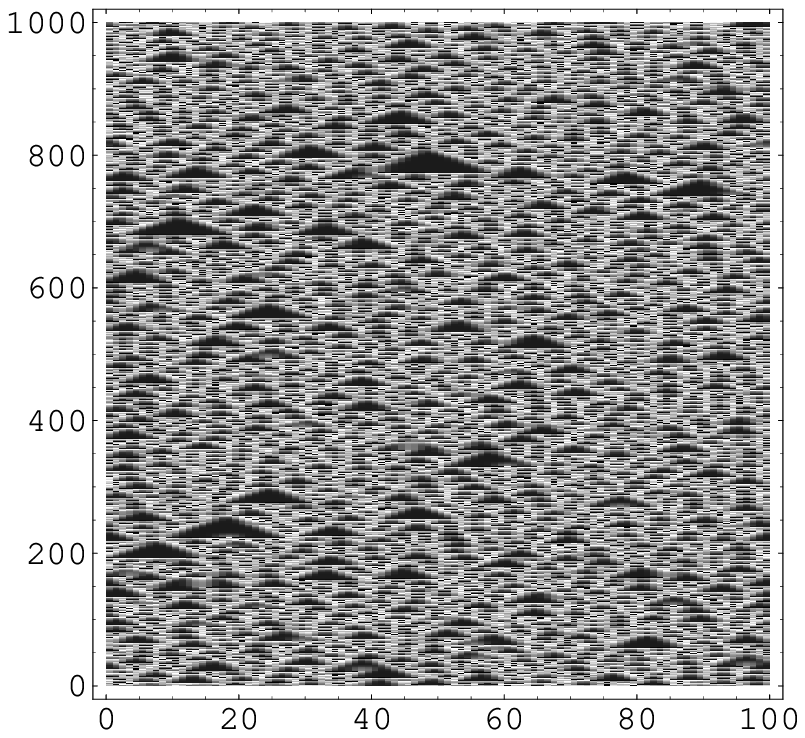}
\caption{Time evolution of the coupled circle map lattice defined
through  eqns. 1-2, in a system of size
$L=100$, for parameters $k=1$ and $\epsilon = 0.63775$. 
Figure~\ref{f.1}(a) is for $\omega = 0.05$ (laminar region) while 
fig. \ref{f.1}(b)
is for $\omega = 0.1$
(turbulent region).  The horizontal axis is the site index $i = 1,
\dots L$; the vertical axis denotes discrete time $t$.}
\label{f.1}
\end{figure}

\begin{figure}
\twoimages[scale=0.875]{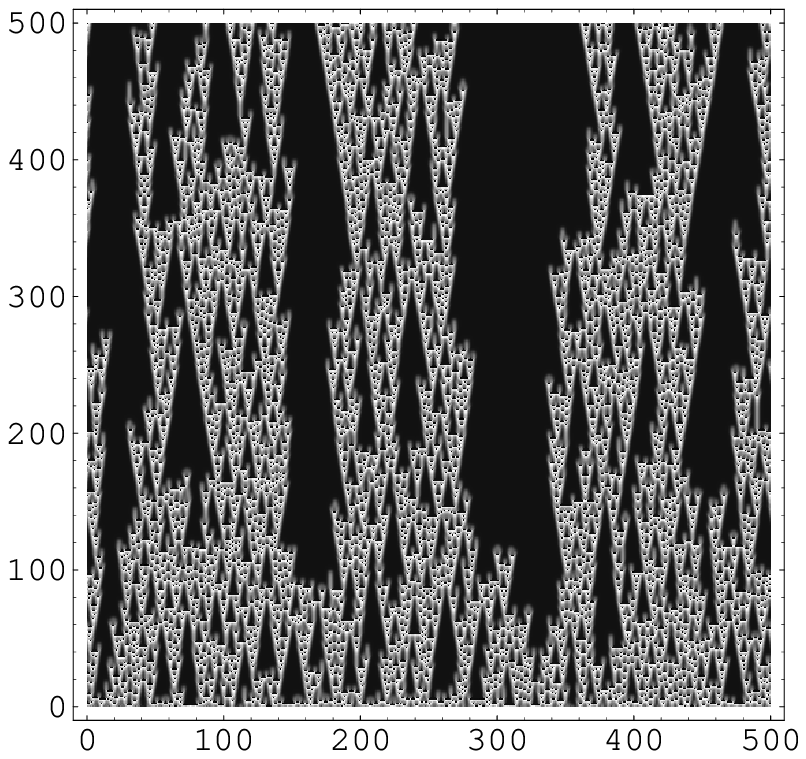}{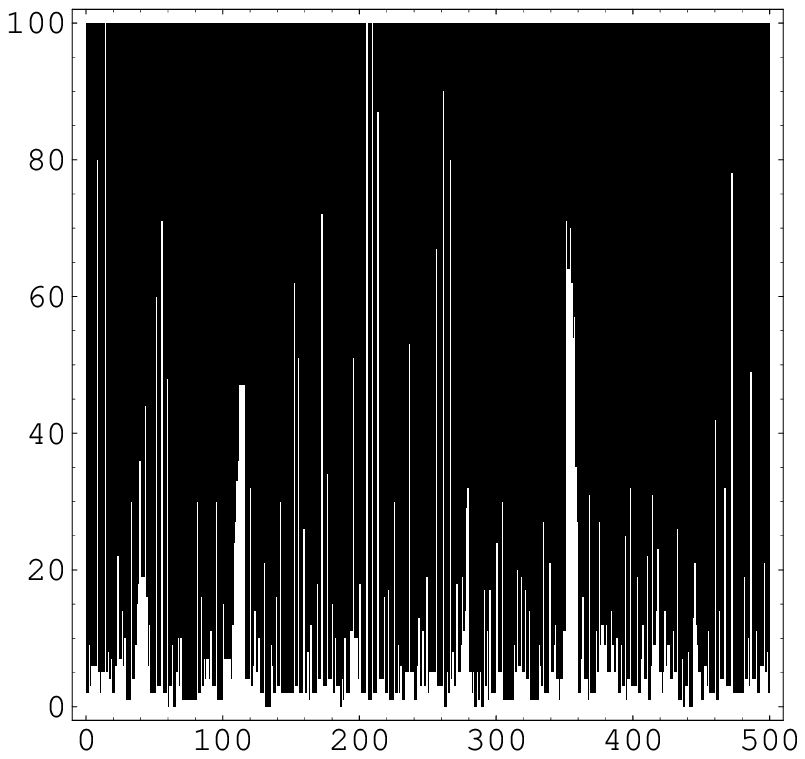}
\caption{Time evolution of the coupled circle map lattice defined
through  
eqns. 1-2,
in a system of size
$L=500$ at the critical point ($k=1$, $\omega = 0.068$, $\epsilon = 0.63775$).
The horizontal axis is the site index
$i = 1, \dots L$ and the vertical axis denotes discrete time $t$. Fig. 2(a)
is obtained from a density plot of the actual $x_{i,t}$ values
(the absorbing regions appear dark). Fig. 2(b) is a plot of
the persistent sites (marked in white) {\it vs.\/} time.}
\label{f.2}
\end{figure}

The synchronized spatiotemporal fixed point $x^{\star}$ corresponds
to the unique absorbing state.
Turbulent sites are those which have
local variables which differ from $x^\star$.
The fixed point solution $x^{\star}$ is easily obtained:
\begin{equation}
\label{three}
x^{\star} = \frac{1}{2
\pi} \sin^{-1} ( \frac{2 \pi \omega}{k} ).
\end{equation}
Figures~\ref{f.1}(a),\ref{f.1}(b), and \ref{f.2}(a) show space-time
density plots of $x_i$ for three choices of
parameter values.  The site index is plotted
along the $x$ axis and time evolves along the
$y$ direction.  Laminar regions are dark in
this representation.  The first of these plots,
fig. 1(a), is for parameter values for which
the final state is laminar, while the
second, fig. 1(b), is for parameter values in the
chaotic regime. Fig. 2(a) marks the transition
between laminar and chaotic regimes.

The critical exponents 
of this coupled map lattice at
the onset of spatio-temporal intermittency
have been obtained numerically; these 
match directed
percolation values to high accuracy\cite{dp_exp}.  This model,
at the critical points studied in this paper,
constitutes the {\em only} known example of
a coupled map lattice with a {\em unique}
absorbing state, whose transition falls
cleanly in the universality class of directed
percolation\cite{dp_exp,examples}.  Finding
direct transitions to a unique absorbing state
from turbulent states in coupled
map lattices is nontrivial; transitions into
spatio-temporally periodic patterns are far
more common.  As a consequence, even though the
past decade has seen considerable activity in
this field, we know of no other examples of such
critical points in the literature\cite{accuracy}.

A further advantage of this model system is 
the absence of any additional special
spatio-temporal structures, a feature which
commonly disrupts the analogy to DP in extended
dynamical systems\cite{bohr}.
Operationally, this means that no
asynchronicity in updates need be introduced here
to destroy ``solitonic'' behaviour \cite{bohr},
Thus, coupled circle
maps are clean model systems for checking DP
universality.  

Persistence in the context of stochastic
processes is defined as the probability $P(t)$
that a stochastically fluctuating variable
has not crossed a threshold value upto time
$t$\cite{pers0}. For a spin system with discrete
states, such as Ising or Potts spins, persistence
is defined in terms of the probability that a
given spin has not flipped out of its initial
state upto time $t$.  A power-law tail to $P(t)$,
{\it i.e.\/}
\begin{equation}
P(t) \sim 1/t^\theta,
\end{equation}
defines the persistence exponent $\theta$, with $P(t)$
averaged over an ensemble of random initial conditions.

What is the appropriate description of persistence in coupled map
lattices?  A natural
generalization of the ideas above
{\em defines} local persistence 
in terms of the probability that a
local state variable $x_{i, t}$ does not cross the fixed point value
$x^{\star}$ up to time $t$.
With this definition, we study persistence
in the coupled map lattice defined through Eqn.~(\ref{one})
{\it via\/}
simulations
on systems of linear size ranging from $L=10$ to $L=10^5$.  We
start with the local variable $x_{i,0}$ distributed uniformly
in the interval $[0,1]$,
update all sites synchronously, and 
measure the normalized fraction of persistent
sites, defined as sites for which $x_{i,t}$ has not {\em crossed} $x^{\star}$
up to time $t$. Thus, sites for which the sign of ($x_{i,t} - x^{\star}$) has
not changed upto time $t$ are persistent at time $t$.

Fig. 2(b) shows the persistent sites
at the critical point corresponding to the configuration of fig. 2(a).
The inset to Fig. 3 shows plots of
$P(t)$ at three different parameter values, illustrating the following
points: (a) $P(t)$ in the laminar region saturates to a
time-independent value $P_\infty$ at large times, with P$_\infty$
becoming smaller and smaller as the transition is approached; (b)
Precisely at the transition between chaotic and laminar regimes,
$P_\infty$ is zero and $P(t)$ decays as a power law, thereby defining
the local persistence exponent $\theta_l$ and finally; (c) In the chaotic
regime $P(t)$ decays exponentially.  These curves are analogous to data
obtained for persistence in the Domany-Kinzel cellular automaton\cite{hinrich},
with the identification of the turbulent phase here with the active
phase of the Domany-Kinzel model and the laminar phase with the inactive one.

Figure 3 (main panel) shows our most accurate determination of the local
persistence probability distribution $P_l(t)$ at a critical point, on
a system of size $L=10^5$.  We
work at the two critical points known to high accuracy $(k =
1.0, \omega = 0.068, \epsilon = 0.63775)$ and $(k = 1.0, \omega =
0.064,\epsilon = 0.73277)$,  averaging over $10^4$ initial conditions. 
To within the accuracy of our
simulations, the values of $\theta_l$ at both these critical points are
the same, consistent with a persistence exponent of
$\theta_l = 1.49 \pm 0.02$\cite{diffusion}.

\begin{figure}
\onefigure[scale=0.450]{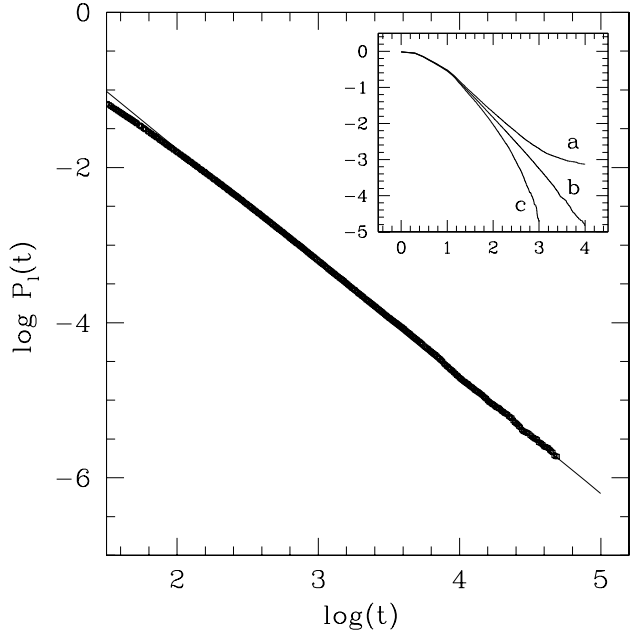}
\caption{  Local persistence distribution $P_l (t)$ vs time $t$ for the 
  critical point at: $k=1$, $\omega = 0.068$, $\epsilon = 0.63775$.
  Inset shows the log-log plot of $P_l (t)$ vs $t$ for parameters
  $k=1$, $\epsilon = 0.63775$ and (a) $\omega = 0.066$ (below critical
  point); (b) $\omega = 0.068$ (at the critical point displaying power
  law scaling) and (c) $\omega = 0.07$ (above critical point).}
\label{f.3}
\end{figure}

If the diverging length (and time) scales at
the continuous directed percolation transition
determine the scaling properties of local
persistence, we expect a scaling form of the type
\begin{equation}
P_l(t,L,\delta) \sim t^{-\theta_l}
{\cal F}_{\pm}(\delta^{\nu_\parallel}t,L^{-z}t).
\end{equation}
Here $z=\nu_\parallel/\nu_\perp$,
with $\nu_\perp$
the standard 1+1 dimensional DP exponent
governing the decay of correlation lengths
in the spatial direction and $\nu_\parallel$
the analogous exponent for the time direction
($\nu_\parallel = 1.733847, \nu_\perp =
1.096854$)\cite{hinrich,dp_experiments}.  ${\cal
F}_{\pm}$ is a universal scaling function 
(the subscript $\pm$ refers to the direction
of approach to the critical point), while $\delta$ measures
the distance from the critical point.  At $\delta
= 0$, ${\cal F}_{\pm}$ behaves asymptotically
as ${\cal F}{\pm}(0,0) \sim~~{\rm const}$; ${\cal
F}_{\pm}(x,0) \sim~~x^{\theta_l}, x \rightarrow
\infty$; ${\cal F}_{\pm}(0,y) =~~{\rm const}, y <<
1$ and ${\cal F}_{\pm}(0,y) \sim~~y^{\theta_l}, y
> 1$.  This scaling function is similar to one
proposed in the study of the spatial scaling
properties of persistence\cite{Ray}.  Fig. 5
shows this scaling function at criticality
$(\delta = 0)$, for various system sizes $L$.
The best data collapse is obtained for $z=1.58$
and $\theta_l = 1.485$. The quality of the
data collapse verifies that the scaling forms
are fully consistent with directed percolation
exponents\cite{hinrich_review,hinrich}.

\begin{figure}
\onefigure[scale=0.43]{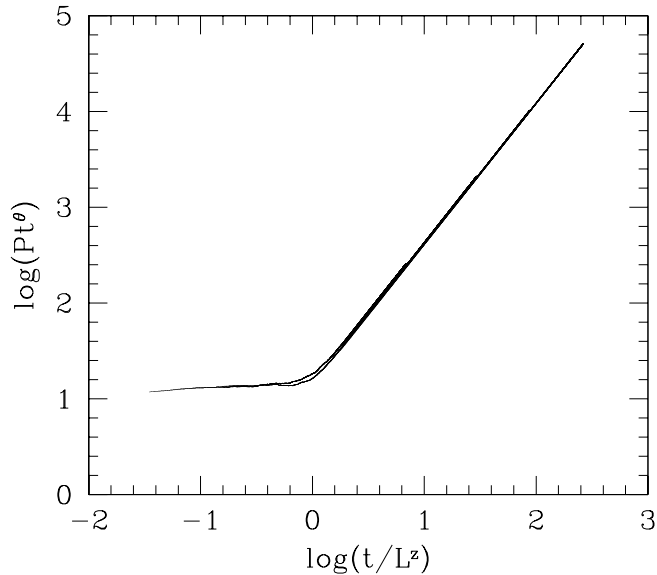}
\caption{ Log-log plot of scaled persistence probability $P_l (t)
  t^{\theta_l}$ versus scaled time $t L^{-z}$ for different lattice
  sizes at the critical point: $k=1$, $\omega = 0.068$, $\epsilon =
  0.63775$. The data collapses onto a single curve 
for $L= 50, 100, 300, 500$ and $1000$.
}
\label{f.5}
\end{figure}

Our result for the local persistence exponent
in coupled circle maps at the laminar-turbulent
transition agrees well with the best estimates
for $\theta$ in the Domany-Kinzel automaton
at the directed percolation transition in 1+1
dimensions, where Hinrichsen and Koduvely found
$\theta = 1.5 \pm 0.02$\cite{hinrich}.  A recent study by
Albano and Mu{\~ n}oz finds $\theta = 1.5 \pm
0.01$ for the contact process at its critical
point in $1+1$ dimensions\cite{albano}, a value
in agreement with our simulations.  These papers
also study {\em global} persistence properties\cite{global_main},
drawing on calculations of the global persistence
exponent $\theta_g$ for phase ordering in
Ising and Potts systems, where such exponents
differ from their local values.  However, the
properties of $\theta_g$ remain controversial
for DP-related problems\cite{global}.
Given the somewhat 
contrived definition of global persistence 
in directed percolation, we believe that
it would certainly be far harder to access this 
exponent experimentally than the local 
exponent.

In summary, this Letter presents suggestive
numerical evidence for the universality of
the persistence exponent at the onset of
spatio-temporal intermittency in coupled map
lattices.  We obtain a persistence exponent
whose value is numerically identical to values
obtained at the directed percolation transition
in 1+1 dimensions in a variety of systems,
including the Domany-Kinzel automaton and
the one-dimensional contact process\cite{sandpiles}.
However, the model we study here is
both {\em deterministic} and possesses a {\em
continuum} degree of freedom at each lattice
site, unlike these other models.

The implications of these results for experiments
is particularly noteworthy.  Spatio-temporal
intermittency is a common phenomena of many
extended systems.  It is seen
in experiments on convection \cite{convect} and
in the ``printers instability'' \cite{print}.
The state variables of coupled map lattices can
often be identified with physical quantities
such as voltages, currents, pressures,
temperatures, concentrations or velocities.
A recent experiment by Rupp,
Richter and Rehberg\cite{rupp} obtains
directed percolation exponents at the
transition to spatio-temporal intermittency in
a one-dimensional system of ferrofluid spikes,
driven by an external oscillating magnetic
field. 
We believe that obtaining the local persistence
exponent from an experiment such as this should
be fairly straightforward.  
We hope this Letter stimulates further work towards
identifying new candidates for measuring
persistence exponents in real world phenomena.




\begin{thebibliography}{0}
\bibitem{type1} This type of spatiotemporal intermittency is 
known as type I spatiotemporal intermittency in the literature.
\bibitem{accuracy} We also point out that the location of the critical
point must be known to high accuracy for a precise evaluation of
$\theta$ and the standard DP exponents.
\end{thebibliography}
\end{document}